\documentclass[pre,aps,twocolumn]{revtex4}
\usepackage{graphicx}

\newcommand{\beq}{\begin{equation}}
\newcommand{\eeq}{\end{equation}}
\newcommand{\ba}{\begin{array}}
\newcommand{\ea}{\end{array}}
\newcommand{\bean}{\begin{eqnarray*}}
\newcommand{\eean}{\end{eqnarray}}
\newcommand{\bea}{\begin{eqnarray}}
\newcommand{\eea}{\end{eqnarray}}
\newcommand{\bc}{\begin{center}}
\newcommand{\ec}{\end{center}}
\newcommand{\bt}{\begin{table}}
\newcommand{\et}{\end{table}}

\newcommand{\beqno}{\begin{displaymath}}
\newcommand{\eeqno}{\end{displaymath}}

\newcommand{\been}{\begin{enumerate}}
\newcommand{\een}{\end{enumerate}}

\begin{document}

\title{Skyrmion-like excitations in dynamical lattices}

\author{P. G.\ Kevrekidis$^{1}$, R.\ Carretero-Gonz\'{a}lez$^{2}$, D. J.\
Frantzeskakis$^{3}$, B. A.\ Malomed$^{4}$, and F. K. Diakonos$^{3}$}
\affiliation{$^{1}$ Department of Mathematics and Statistics, University of
Massachusetts, Amherst MA 01003-4515\\
$^2$ Nonlinear Dynamical Systems Group, San Diego State University, San
Diego CA, 92182-7720\\
$^3$ Department of Physics, University of Athens, Panepistimiopolis,
Zografos, Athens 15784, Greece \\
$^4$ Department of Interdisciplinary Studies, Faculty of Engineering, Tel
Aviv University, Tel Aviv 69978, Israel}

\begin{abstract}
We construct discrete analogs of Skyrmions in nonlinear dynamical lattices. 
The Skyrmion is built as a vortex soliton of a complex field, coupled to a dark
radial soliton of a real field. Adjusting the Skyrmion ansatz to the lattice setting
allows us to construct a \textit{baby-Skyrmion} in two dimensions
(2D) and extend it into the 3D case (1D counterparts of the Skyrmions are also found). 
Stability limits for these patterns are obtained analytically and verified numerically. 
The dynamics of unstable discrete Skyrmions is explored, and their stabilization 
by external potentials is discussed.
\end{abstract}

\maketitle

\textit{Introduction}. 
Recently, studies of intrinsic localized
modes (ILMs) in nonlinear lattice systems have drawn much
attention \cite{review} due to their relevance 
to various physical problems
including, \textit{inter alia}, optical waveguide arrays
\cite{mora}, photonic crystals \cite{PhotCryst}, Bose-Einstein
condensates (BECs) trapped in deep optical lattices (OLs)
\cite{tromb}, and Josephson-junction ladders \cite{alex}.

A wide variety of self-supporting ILM type of excitations have
been predicted theoretically and observed experimentally.
Prominent examples are bright and dark optical discrete solitons
\cite{6,7} in AlGaAs waveguide arrays \cite{7}, multi-dimensional
solitons in photonic lattices 
\cite{8,9,10}, 
discrete vortex solitons
\cite{10a,11,12}, lattice dipole solitons \cite{13},
multi-component solitons \cite{13a}, soliton trains \cite{14}, 
necklace solitons \cite{14a}, and so on. Parallel to these achievements
in optics, a remarkable recent result was the creation of
gap solitons in BECs\ loaded in OLs \cite{gap}. In fact, BECs offer
an interesting implementation of nonlinear dynamics in 
\textit{confined lattices}, due to the fact that experiments are
always run in an external trap \cite{tromb}. 
These developments raise the question whether counterparts of more
complex structures, that were originally proposed in continuum
media within field-theoretical contexts, can be predicted and
observed in dynamical lattices. 

Among the most fundamental objects of this type are the three-dimensional (3D) Skyrmions, 
initially proposed to explain topologically the origin of the baryon quantum number 
\cite{skyrme}. On the other hand, their two-dimensional (2D) version, so-called ``baby-Skyrmions'',  
were used to model bubble generation in condensed-matter systems in the presence of an   
external magnetic field. In particular, baby-Skyrmions 
seem to have a central role in the dissapearance of antiferromagnetism and the onset of high-$T_c$ 
superconductivity \cite{hightc}, as well as in the ground state properties of the 
quantum Hall ferromagnets \cite{qhf}. Recently, stable Skyrmions 
have also been predicted in BECs \cite{becskyrme}. 
On the other hand, as concerns {\it discrete} systems, stable discrete Skyrmions on lattices were considered 
in the framework of the Heisenberg model for magnetism \cite{ward},
as well as in the 2D Hubbard model \cite{seibold}.
Additionally, Skyrme lattices are also very interesting 
due to their ability to describe electron spin 
textures in quantum Hall systems \cite{elspin}. Importantly, 
discrete Skyrmions play also an 
essential role for the quantization of the original Skyrme model \cite{schramm}, the  
continuum version of which is non-renormalizable. 

In the present work, we discuss the discretization of Skyrmions from a different point of view. 
Putting the continuity aspect on second priority, we develop discrete solitons with 
topological properties equivalent to those of Skyrmions in the continuum field theory. 
We show that the proposed Skyrmion-like structures (for simplicity called ``Skyrmions'' hereafter) 
can be stabilized on the dynamical lattice under consideration, namely a 
nonlinear Schr{\"{o}}dinger lattice. 

We aim to construct discrete Skyrmion states in a paradigmatic nonlinear-lattice model, 
\textit{viz}., the discrete nonlinear Schr{\"{o}}dinger (DNLS) equation. 
DNLS 
is a universal envelope wave equation for a variety of Hamiltonian systems 
(such as, e.g., nonlinear Klein-Gordon lattice models), and, moreover, is a direct model for 
BECs trapped in strong OLs \cite{tromb} and crystals built of microresonators \cite{photons}; 
in addition, its 2D version models optical waveguide arrays \cite{mora,6}. 
As Skyrmions necessarily involve (through the so-called 
{\it hedgehog ansatz} \cite{weidig,piette}) \emph{three} scalar fields, 
and are characterized by two independent topological charges, their 
description requires two complex field variables. Thus, in this work we study a
two-component DNLS equation. This model is directly
relevant to waveguide arrays \cite{13a}, when the light has
different polarizations or frequencies, and to binary BECs 
composed as spin state mixtures 
of the same isotope 
\cite{hall}.

The paper is structured as follows: In the next section, we present the
model and some analytical considerations. Then, we report numerical results
for 1D, 2D, and 3D lattices, and, finally, we summarize our findings. 

\textit{Model and Analytics}. We consider a vector DNLS
equation for $\mathbf{\phi }\equiv (\phi _{1},\phi _{2})^{T}$ on the
cubic/square lattice,
\begin{equation}
i\dot{{\mathbf{\phi }}}_{\mathbf{n}}=-C\Delta ^{(d)}{\mathbf{\phi
}}_{\mathbf{n}}+({\mathbf{\phi }}^{T}G\mathbf{\phi })\mathbf{\phi
},\quad G=\left(
\begin{array}{cc}
1 & 0 \\
0 & \beta
\end{array}\right) ,  \label{NLS}
\end{equation}
where overdot stands for derivative in time (or, in the optical-waveguide model, in the propagation
distance), $C$ is a coupling constant, the parameter $\beta$ characterizes the intra-species interaction, 
and the $D$-dimensional discrete Laplacian is 
$\Delta^{(D)}{\mathbf{\phi }}_{\mathbf{n}}\equiv \sum_{\mathbf{m}=<n>}{\mathbf{\phi }}_{\mathbf{m}}-2D{\mathbf{\phi
}}_{\mathbf{n}}$, with $<n>$ standing for the nearest-neighbor
shell of $\mathbf{n}$ [the latter is a vector index in the
$D$-dimensional case --- e.g., $\mathbf{n}=\left(
n_{1},n_{2},n_{3}\right) $ in 3D]. Stationary solutions are looked
for as $\mathbf{\phi }_{\mathbf{n}}=\exp (-i\Lambda
t)(u_{\mathbf{n}},v_{\mathbf{n}})^{T}$, where $\Lambda $ is the frequency
(or chemical potential in the context of BECs), and 
$u_{\mathbf{n}},v_{\mathbf{n}}$ obey the following equations
\begin{eqnarray}
\Lambda u_{\mathbf{n}} &=&-C\Delta u_{\mathbf{n}}+
(|u_{\mathbf{n}}|^{2}+\beta |v_{\mathbf{n}}|^{2})u_{\mathbf{n}}, \\
\Lambda v_{\mathbf{n}} &=&-C\Delta
v_{\mathbf{n}}+(|v_{\mathbf{n}}|^{2}+\beta
|u_{\mathbf{n}}|^{2})v_{\mathbf{n}}.  
\label{standing}
\end{eqnarray}
Note that the nonlinear-interaction matrix $G$ in Eq. (\ref{NLS}) implies
self-repulsion of the lattice field, while the model with
self-attraction, i.e., $G$ replaced by $-G$, can be transformed
into the present form by the known \textit{staggering}
\textit{transformation}; for instance, $\left(
u_{\mathbf{n}},v_{\mathbf{n}}\right) \equiv
(-1)^{n_{1}+n_{2}+n_{3}}\left(\tilde{u}_{\mathbf{n}},\tilde{v}_{\mathbf{n}}\right) $ in the 3D case.

We first construct solutions with the desired structures in 
the so-called anti-continuum (AC) limit of $C=0$, and then continue
solution branches to $C>0$, by means of a fixed-point iteration
converging to the relevant solutions. Once these have been
obtained, linear-stability analysis is performed for a perturbed
solution, $\mathbf{\phi }_{\mathbf{n}}^{(\mathrm{pert})}=\left(
\mathbf{\phi }_{\mathbf{n}}+\mathbf{a}_{\mathbf{n}}e^{\lambda
t}+\mathbf{b}_{\mathbf{n}}e^{\lambda ^{\ast }t}\right)
e^{-i\Lambda t}$, where $\mathbf{a}_{\mathbf{n}}$ and
$\mathbf{b}_{\mathbf{n}}$ constitute an eigenmode of infinitesimal
perturbations, and $\lambda $ is the
corresponding eigenvalue. 
The stationary solution is unstable if at least one pair of
$\lambda $ has nonzero real part.

We start 
with the 2D case, where the 
baby-Skyrmion can be found in the continuum model, using a ``hedgehog" ansatz (in the polar coordinates) 
\cite{weidig,piette}, namely 
\begin{equation}
\mathbf{\Psi}=\left[ \sin (f(r))\cos (k\theta ),~\sin (f(r))\sin
(k\theta ),~\cos (f(r))\right] ^{T},  
\label{Psi}
\end{equation}
subject to the boundary conditions $\lim_{r\rightarrow 0}f(r)=M\pi
$, and $\lim_{r\rightarrow \infty }f(r)=N\pi$, with integer $k$,
$M$ and $N$. The winding number of the corresponding state is
$[(-1)^{N}-(-1)^{M}]k/2$. We will try to construct a counterpart 
of ansatz (\ref{Psi}) on the 2D lattice. In particular, combining 
the first two components of ansatz (\ref{Psi}) as $\Psi _{1}+i\Psi_{2}$, 
we obtain a complex field with vorticity $k$. To
consider the fundamental 
Skyrmions, we set
$k=1$ and $M=0$, $N=1$. Accordingly, the complex field is a
localized discrete vortex, while the remaining field may be real,
taking the form of a dark soliton in the (quasi-)radial direction
on the lattice.  For $C=0$, one can construct a
discrete vortex (alias \textit{vortex cross}) by assigning the
complex fields phases $0,\pi /2,\pi ,3\pi /2$ and modulus
$\sqrt{\Lambda }$ at four sites, $(m,n)=(1,0)$, $(0,1)$, $(-1,0)$
and $(0,-1)$, which surround the central point \cite{10a}. The
radial dark soliton in the real field is achieved, in the same AC
limit, by setting $v_{0,0}=1$ and $v_{m,n}=-1$ at all other sites,
with the exception of the above-mentioned four sites surrounding
the origin, where $v_{m,n}=0$. Notice that this corresponds to a
hedgehog ansatz with $f(0)=0 $, $f(1)=\pi /2$ and $f(r>1)=\pi$. 
Of course, the size of the vortex structure may be made larger 
in the AC limit, but this adversely affects 
the stability of the vortex \cite{pelin}, 
therefore we do not examine such cases here.

We note that the above 2D structure suggests a 1D analog of the
Skyrmion, essentially as a cross-section of the 2D profile. In
particular, in the AC limit, the 1D configuration becomes
$u_{n}=\sqrt{\Lambda }(\delta _{n,1}-\delta _{n,-1})$,
$v_{0}=\sqrt{\Lambda }$, $v_{1}=v_{-1}=0$ and
$v_{n}=-\sqrt{\Lambda }$ for $|n|>1$. This 1D structure
essentially consists of a twisted localized mode \cite{tlm} in the
one field, coupled to a pair of discrete dark solitons in the other. 
A variety of generalizations to the 3D case, resulting from Skyrmion's rotation, are possible, 
but we limit our considerations to the simplest setting where
$u_{\mathbf{n}}$ is a planar excitation (as above) with the
well-defined vorticity around the third dimension, while
$v_{0,0,0}=\sqrt{\Lambda }$, the field vanishes at the immediate
neighbors of $(0,0,0)$ and has a value of $-\sqrt{\Lambda }$ elsewhere.

To examine instabilities relevant to the discrete Skyrmion, we first examine
the dispersion relation for the equations linearized around such a solution
(i.e., the continuous spectrum). While the full spectrum also contains
isolated eigenvalues, it is its continuous component which is primarily
responsible for instabilities (see below). Thus, we perturb the asymptotic
lattice field, far from the Skyrmion's center, as follows:
\begin{eqnarray}
\phi _{1} &=&\epsilon Ae^{i(\omega t+kn)}+\epsilon Be^{-i(\omega ^{\star
}t+kn)},  \label{lin1} \\
\phi _{2} &=&-1+\epsilon Ce^{i(\omega t+kn)}+\epsilon De^{-i(\omega ^{\star
}t+kn)}.  \label{lin2}
\end{eqnarray}
Using Eqs. (\ref{lin1})-(\ref{lin2}) in Eq. (\ref{NLS}) yields two excitation branches,
associated, respectively, with the zero and non-vanishing backgrounds of the
first and second fields,
\begin{eqnarray}
\omega =\pm \left[ \Lambda -\beta -4CD\sin ^{2}({k}/{2})\right], \\
\omega =\pm \sqrt{\Lambda +4CD\sin ^{2}({k}/{2})-\Lambda ^{2}},  
\label{lin4}
\end{eqnarray}
which restricts the relevant spectral bands to 
$\pm [\Lambda-\beta-4CD,$ $\Lambda -\beta ]$
and $[-\sqrt{(\Lambda +4CD)^{2}-\Lambda^{2}},\sqrt{(\Lambda +4CD)^{2}-\Lambda ^{2}}]$. 
These bands have opposite \textit{Krein signatures}
\cite{krein}, hence their collision, occuring with the
increase of $C$, at
\begin{equation}
C_{\mathrm{cr}}^{(D)}={(\beta -\Lambda )^{2}}/[{8D(2\Lambda -\beta )}],
\label{prediction}
\end{equation}
will generate complex eigenvalues, i.e., instability. 
Thus, 
the discrete Skyrmions may only be stable in 
the interval of $0\leq C<C_{\mathrm{cr}}^{(D)}$; in particular,
for $C_{\mathrm{cr}}^{(D)}>$ $0$ one requires $2\Lambda >\beta $.
It is interesting to mention that $C_{cr}^{(D)}\rightarrow \infty
$ (the continuum limit) as $\Lambda -\beta /2\rightarrow +0$;
however, analysis of this special case is beyond the scope of this
paper.

\begin{figure}[tbp]
\begin{center}
\hskip-0.15cm
\begin{tabular}{cc}
\includegraphics[width=2.75cm]{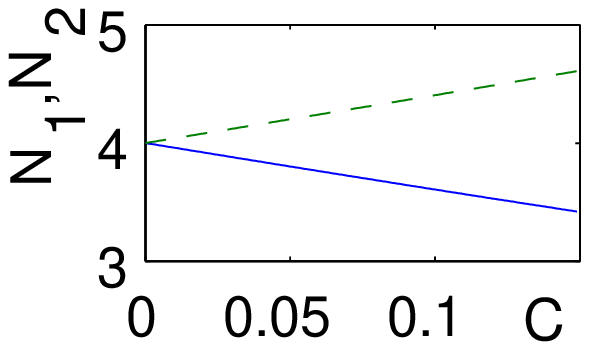}
\includegraphics[width=2.75cm]{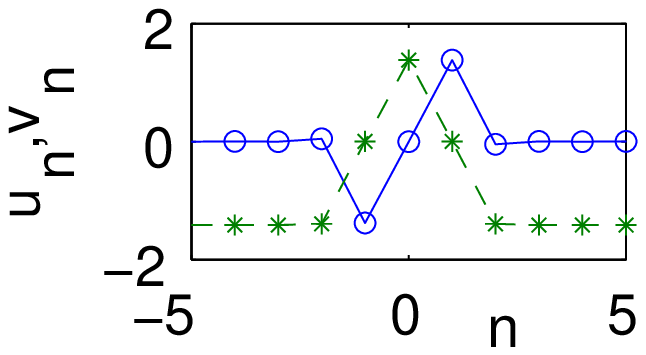}
\includegraphics[width=2.75cm]{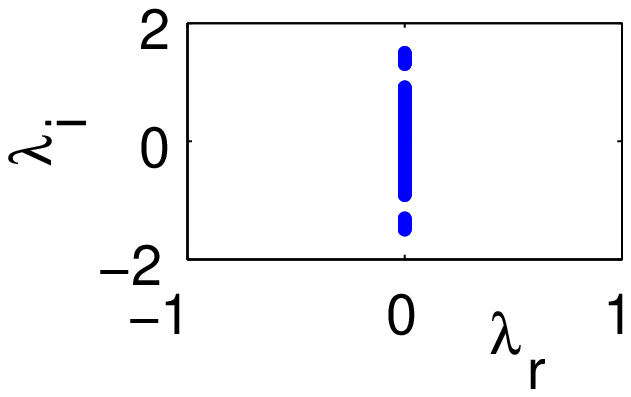} &  \\[-1.0ex]
~\includegraphics[width=2.75cm]{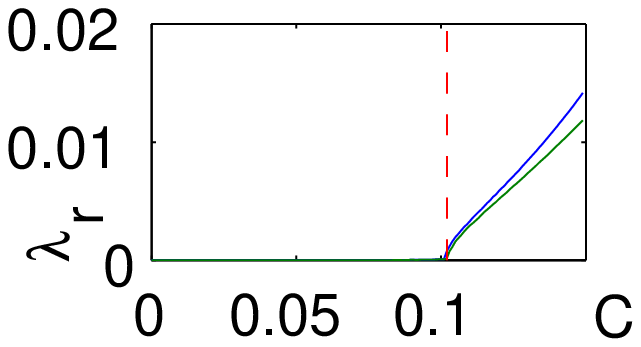}
\includegraphics[width=2.75cm]{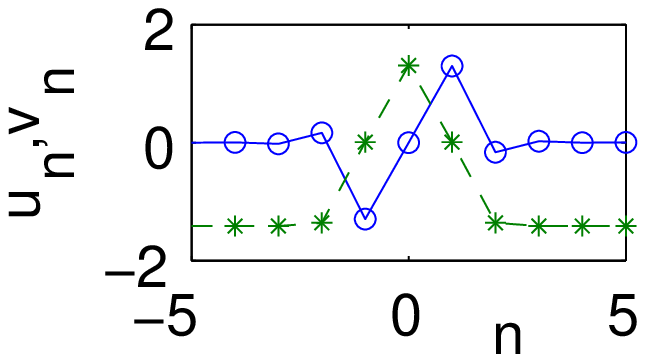}
\includegraphics[width=2.75cm]{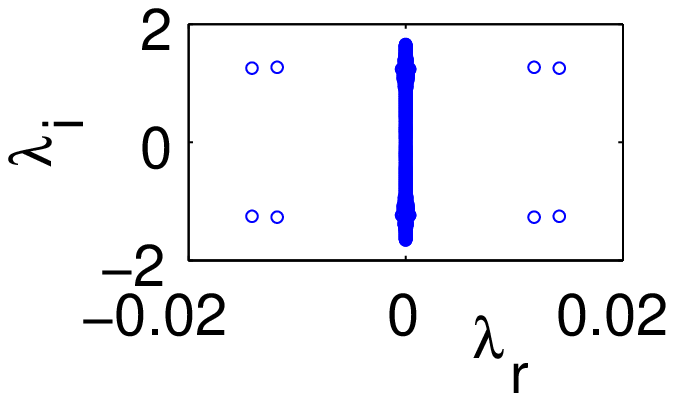} &
\end{tabular}\end{center}
\par
\vskip-0.7cm \caption{(Color online) 
1D discrete analog of the Skyrmion. Top and bottom panels in the left column
show, respectively, norms $N_{1},N_{2}$ (defined in the text) and
two most unstable eigenmodes vs.~coupling $C$ [the vertical dashed
line is the instability onset as predicted by Eq. (\protect
\ref{prediction})]. Upper and bottom panels in the middle and
right columns display, respectively, the solution for $C=0.05$ and
$C=0.15$, and the corresponding spectral planes $(\protect\lambda
_{r},\protect\lambda _{i})$ of the eigenvalue $\protect\lambda
=\protect\lambda _{r}+i\protect\lambda _{i}$.}
\label{fig1}
\end{figure}

\textit{Numerical Results}. We first display numerical findings
for the 1D case in Fig.~\ref{fig1}. Shown are norms of the
solution, $N_{1}=\sum_{\mathbf{n}}|u_{\mathbf{n}}|^{2}$ and
$N_{2}=\sum_{\mathbf{n}}(\Lambda -|v_{\mathbf{n}}|^{2})$, and most
unstable eigenmodes and eigenvalues (computed on a lattice with
400 sites). In this case, Eq.\ (\ref{prediction}) predicts
$C_{\mathrm{cr}}^{(1)}\approx 0.102083$, which coincides with the
numerical finding ($0.102\pm 0.0005$). Examples of stable and
unstable 1D Skyrmions are included too, for $C=0.05$ and $C=0.15$.
As expected (see above), the destabilization occurs, indeed, via
the collision of two continuous bands of eigenvalues at the
critical point. It has been checked that the numerical results,
displayed here for $\Lambda =2=8\beta $, adequately represent a
large area in the parameter space.

\begin{figure}[tbp]
\begin{center}
\begin{tabular}{ll}
\hskip-0.2cm\includegraphics[width=5.1cm]{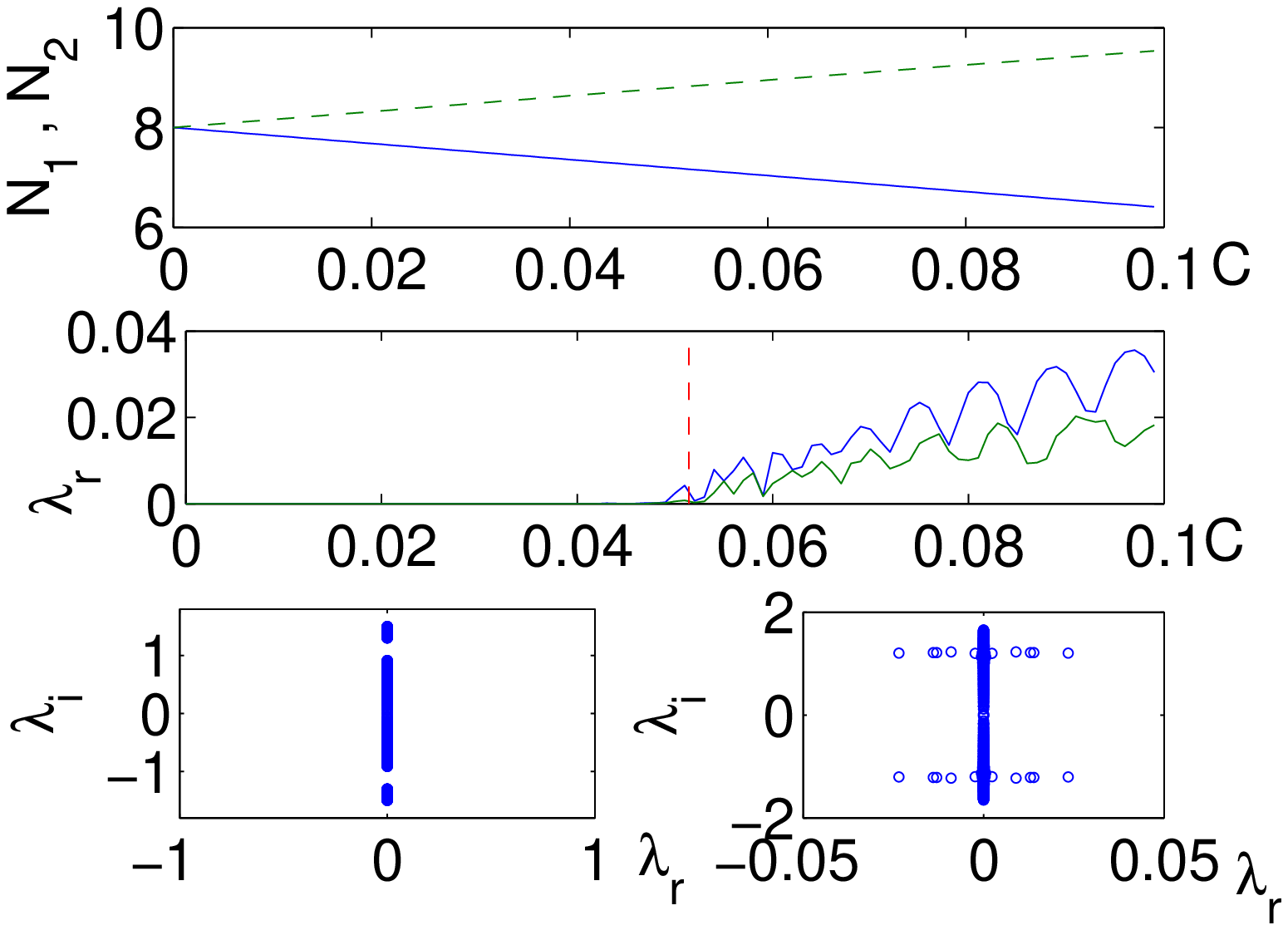} \hskip0.5cm & \hskip0.0cm
\includegraphics[width=3cm,height=3.7cm]{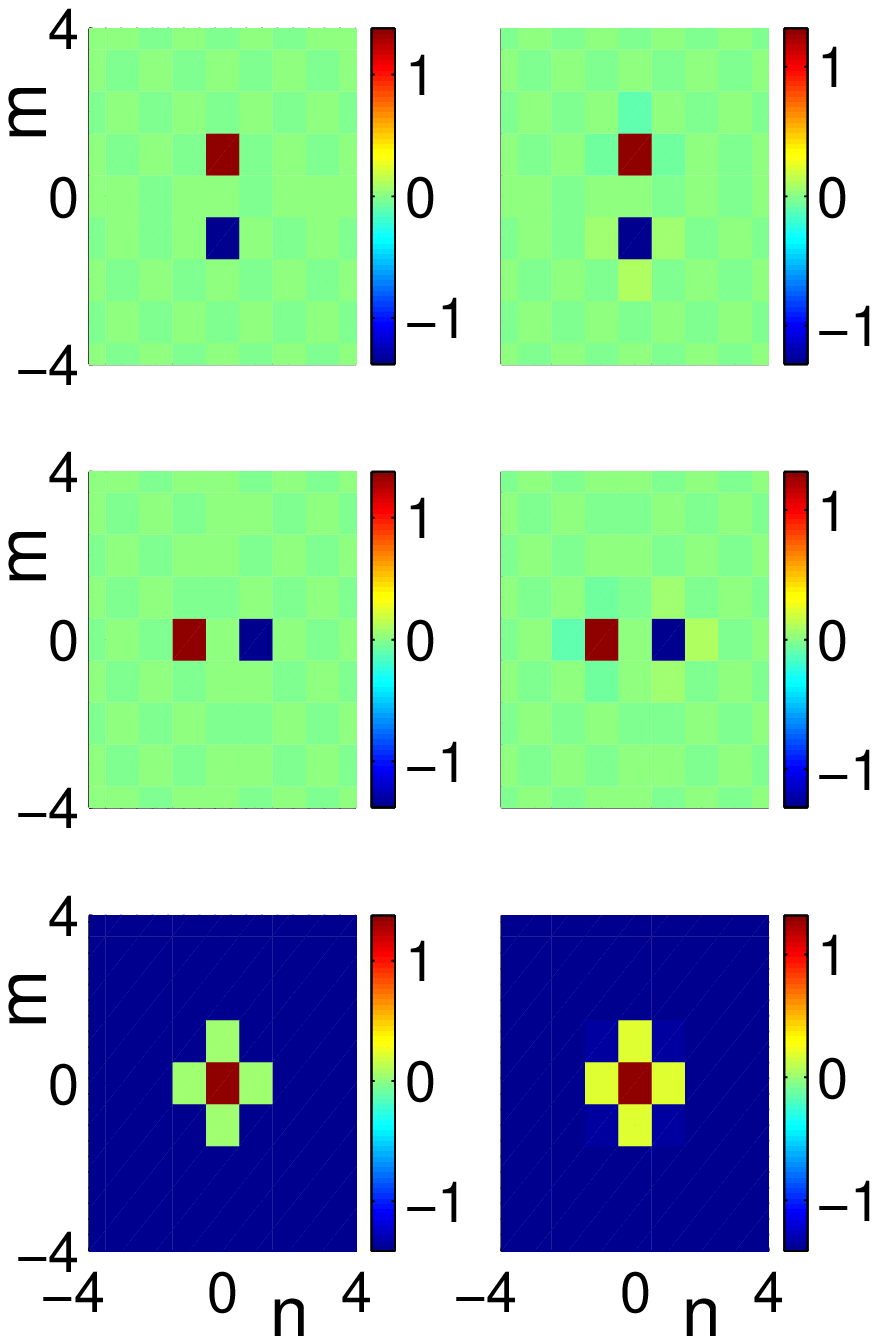} \\[-4.0ex]
&
\end{tabular}\end{center}
\caption{(Color online) Same diagnostics as in
Fig.~\protect\ref{fig1} for 
discrete 2D (\textit{baby}-) Skyrmions. Left and right paired panels pertain,
respectively, to $C=0.025$ and $C=0.075$. Solution profiles are
shown on the right, with the top, middle, and bottom rows
displaying countours of real and imaginary parts of the first
(complex) field, and the second (real) field.}
\label{fig2}
\end{figure}

Results for the 2D case are presented in Fig.~\ref{fig2}. We again
observe excellent agreement of the theoretical prediction
(\ref{prediction}) for the instability onset, at
$C=C_{\mathrm{cr}}^{(2)}=0.051042$. Typical solution examples are
shown for $C=0.025$ (stable) and $C=0.075$ (unstable).

\begin{figure}[tbp]
\begin{center}
\hskip-0.2cm\includegraphics[width=3.5cm]{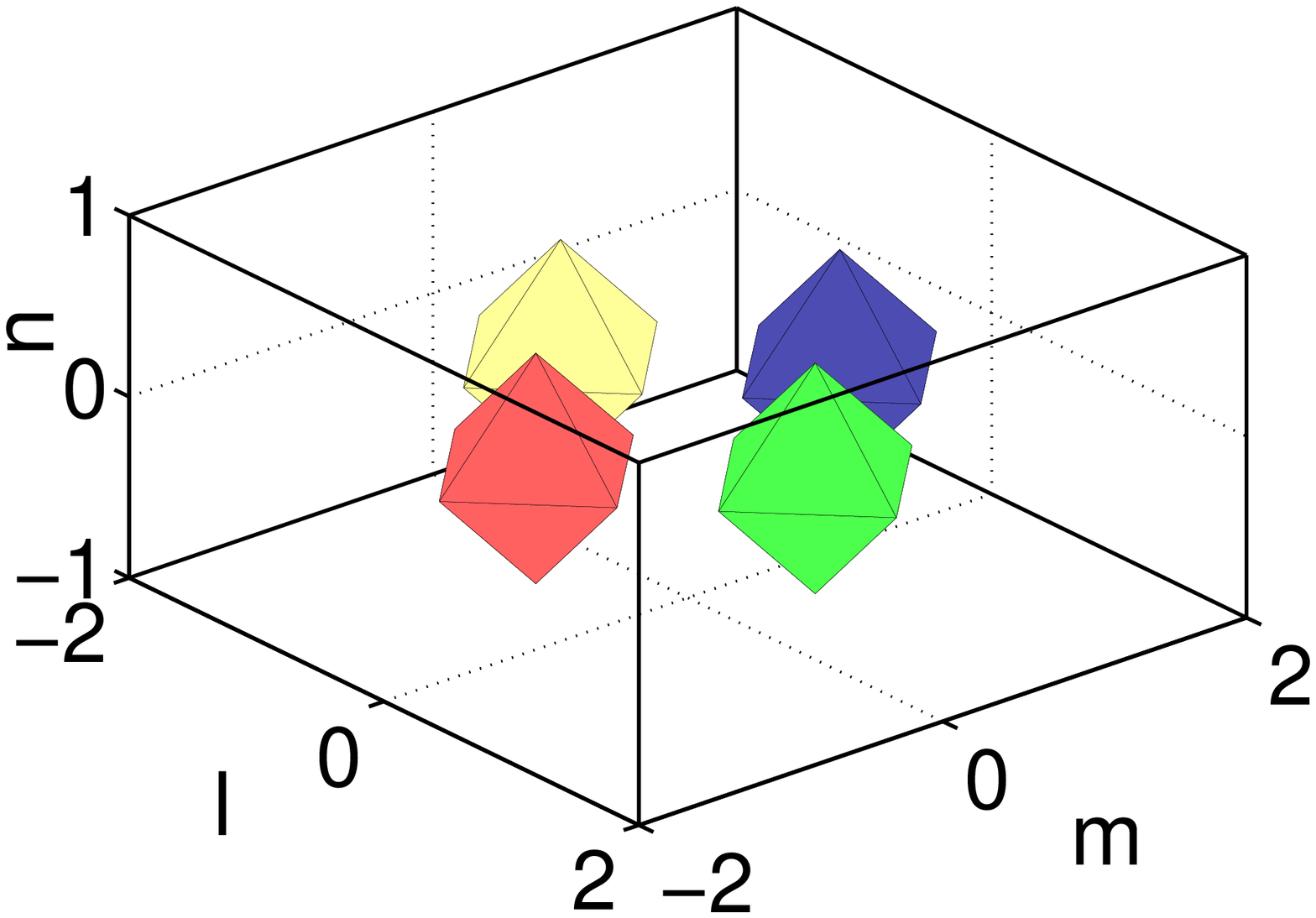}
 ~~
\includegraphics[width=3.5cm]{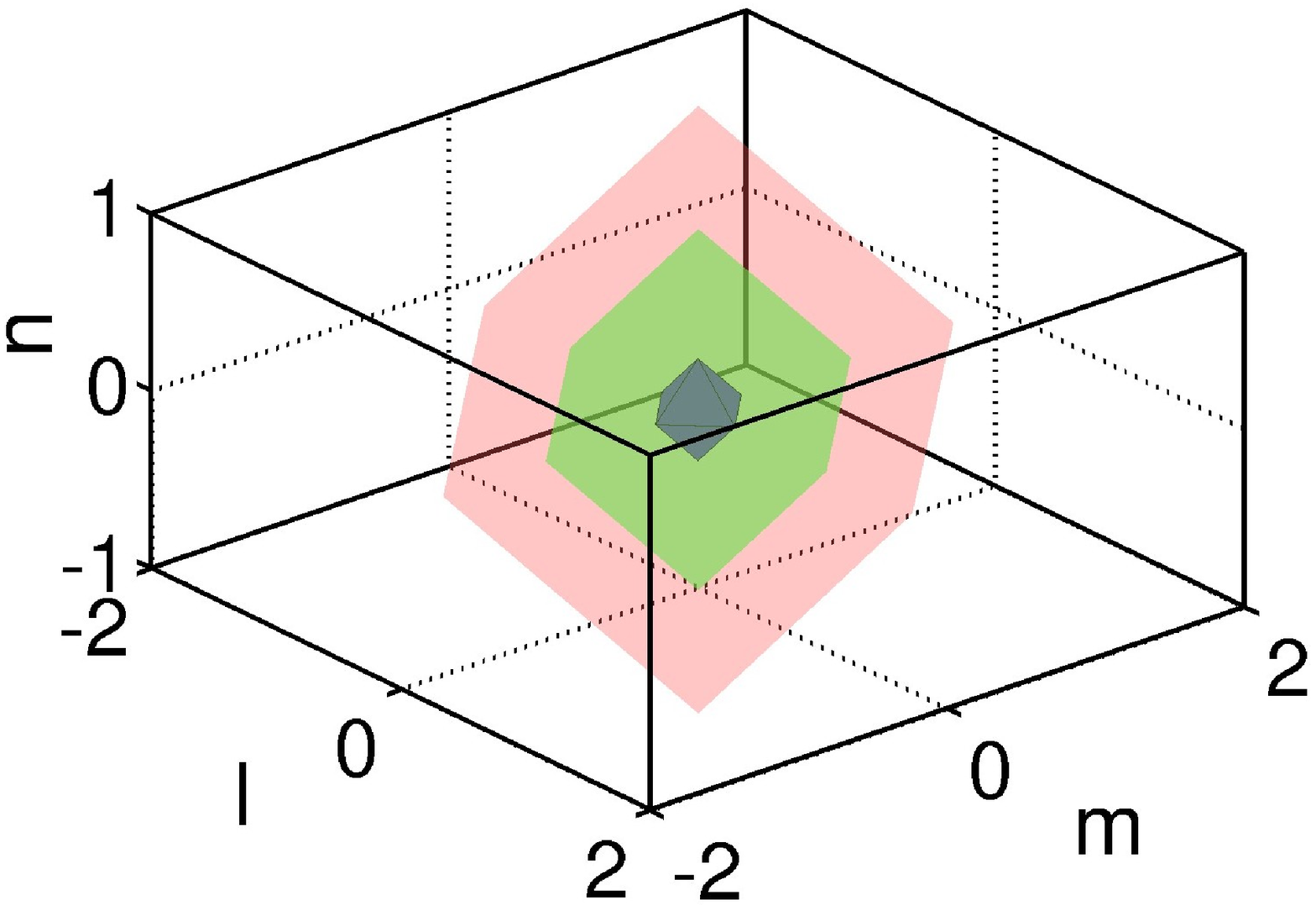}
\end{center}
\par
\vskip-0.5cm \caption{(Color online) A 
3D discrete Skyrmion for $C=0.05$. The left and right panels show,
respectively, contours of the first (complex) field, at
$\mathrm{Re}(\protect\phi _{1})=\pm 1$ (blue/red) and
$\mathrm{Im}(\protect\phi _{1})=\pm 1$ (green/yellow), and of the
second (real) field, at $\protect\phi _{2}=(-1,0,1)$ (red, green
and blue, respectively).}
\label{fig3}
\end{figure}

An example of one of the possible (as mentioned above) 3D
generalizations of the 2D discrete baby-Skyrmion is shown in Fig.
\ref{fig3}. In this case, the complex field is arranged as a 3D
soliton carrying a vortex in the horizontal plane (cf. Ref.
\cite{we}), while the real component features a 3D radial dark
soliton. A stable 3D discrete Skyrmion is shown for $C=0.01$ (in this case,
$C_{\mathrm{cr}}^{(3)}=0.034028$).

\begin{figure}[tbp]
\begin{center}
\begin{tabular}{ll}
\hskip-0.2cm\includegraphics[width=4.1cm]{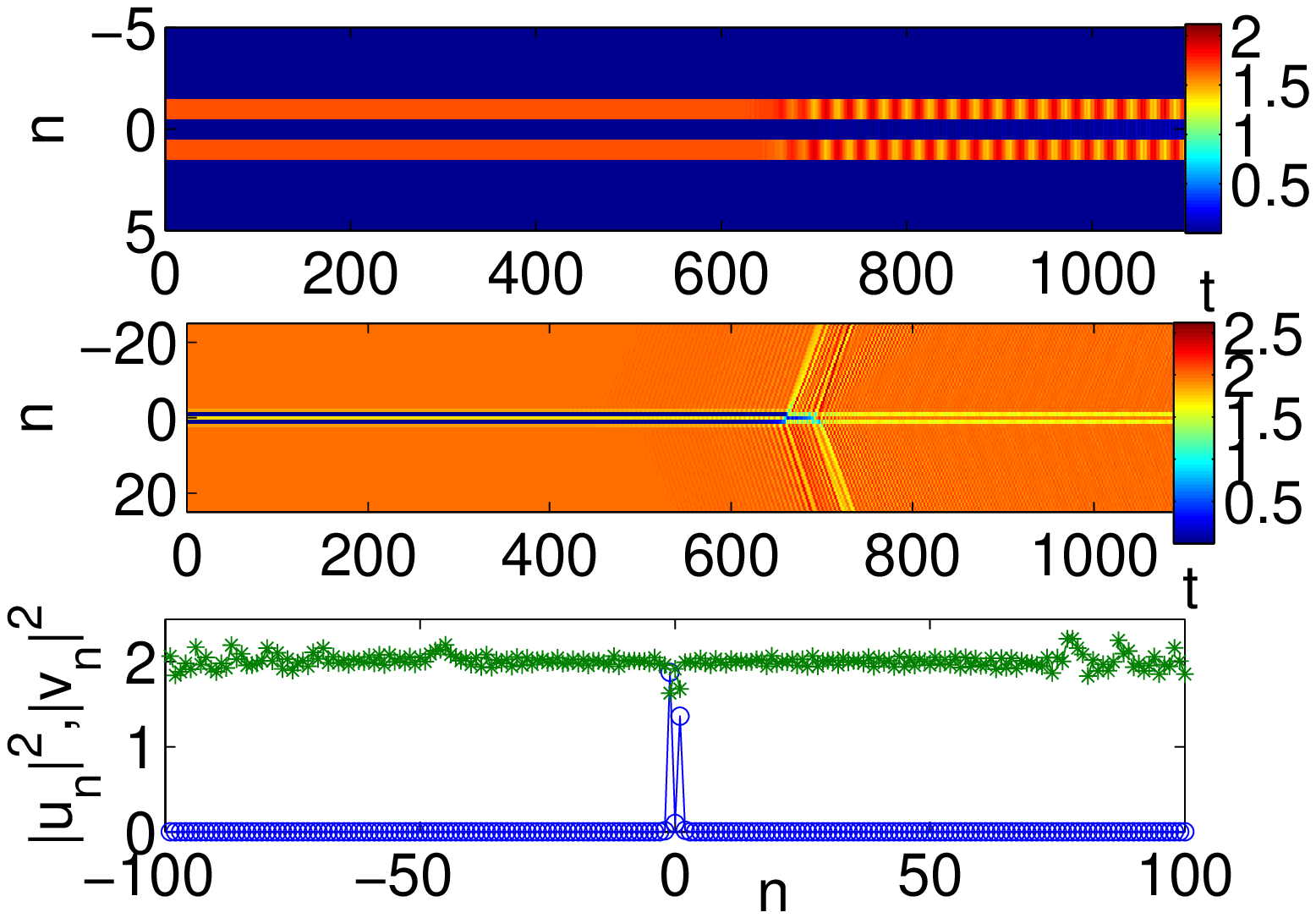} & ~~\hskip0.0cm
\includegraphics[width=4.1cm]{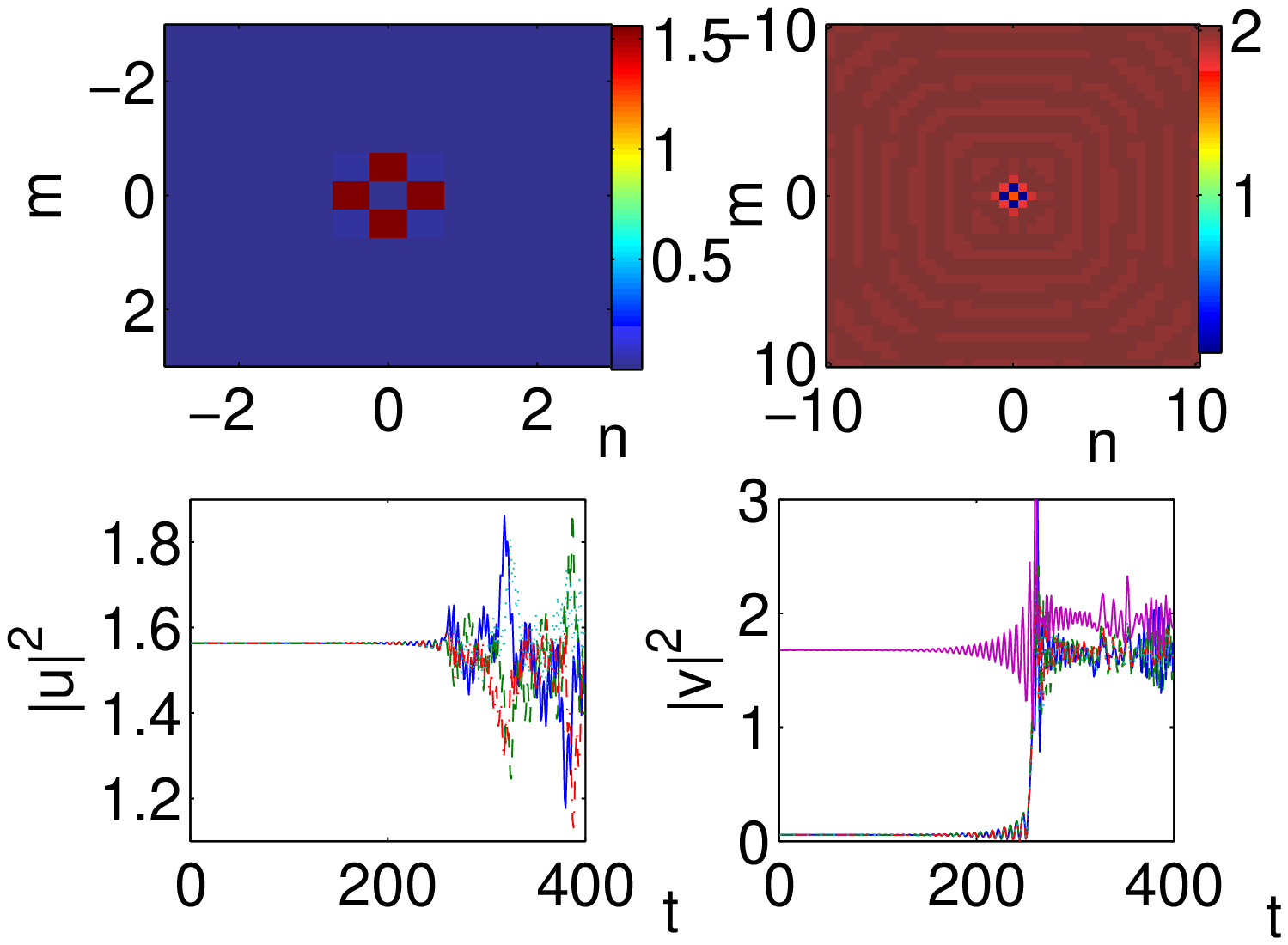} \\[-4.0ex]
&
\end{tabular}\end{center}
\caption{(Color online) Instability development in one- and
two-dimensional discrete Skyrmions. The top, middle, and bottom
left panels show, respectively, space-time contours of the
absolute value of the first (complex) field and of the second
(real) one, and the spatial distribution of the fields at $t=1100$
in the 1D case. The top right panels display the contours of the
    square modulus of the two fields for the 2D case,
at the instability onset, $t=220$, while the panels
beneath them show the evolution of the fields at the central site
and its four neighbors, for the same case.}
\label{fig4}
\end{figure}

The next step is to simulate the evolution of unstable solutions.
As, beyond the primary instability threshold (\ref{prediction}), a
cascade of secondary instabilities is produced by additional
collisions between the continuous spectral bands, one may expect
that the corresponding multitude of unstable eigenmodes leads to a
kind of lattice turbulence, especially because the unstable
eigenmodes are delocalized, as they belong to the continuous
spectrum. Indeed, this is what we observe in direct simulations,
as shown in Fig.~\ref{fig4} for $C=0.149>C_{\mathrm{cr}}^{(1)}$
and $C=0.099>C_{\mathrm{cr}}^{(2)}$ in the 1D and 2D cases.
In the 1D case (the left part of the figure), the weakly unstable
configuration remains undisturbed for a long time, but eventually,
around $t\approx 650$, the instability generates spatial chaos in
the real (second) field component, and breathing in its complex
counterpart. This dynamics persists for long evolution times.
Transition to
chaotic behavior is also observed (in the right part of Fig.
\ref{fig4}) in the 2D case, after the onset of the instability
around $t=220$.

\begin{figure}[t]
\begin{center}
\hskip-0.2cm\includegraphics[width=7.0cm,height=4.5cm]{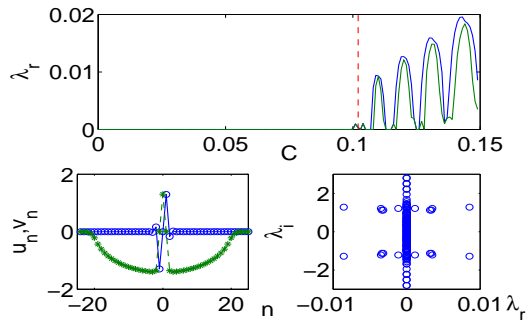}
\end{center}
\par
\vskip-0.75cm \caption{(Color online) The most unstable eigenvalue
(top) for the 1D lattice Skyrmion, and an example of the field
configuration and its stability for $C=0.15$ (bottom left and
right) in the presence of the confining parabolic potential with
$\Omega =0.1$.}
\label{fig5}
\end{figure}

Finally, we also considered the influence of external potentials,
which is necessary in the application to BECs, that 
are typically confined by a parabolic trap. In the 1D case, the latter
amounts to adding a term $(\Omega
^{2}/2)n^{2}\mathbf{\phi }$ to Eq. (\ref{NLS}) (in the 2D and 3D
cases, the trap produces qualitatively similar effects). As seen
in Fig.~\ref{fig5}, the profile of the $v$ field now has a finite
size, as per the Thomas-Fermi approximation \cite{tromb}. The main novel
feature induced by the trap is the appearance of \emph{gaps} in the
linearization spectrum, which leads to the dependence of the
largest unstable eigenvalue on $C$, as shown in the top panel of
Fig.~\ref{fig5}. Interestingly, while the envelope of this
dependence traces a curve similar to that in Fig.~\ref{fig1}, the
gaps lead to \emph{restabilization} of the discrete Skyrmion in
certain intervals. For instance, the solution shown is stable for
$0.113\leq C\leq 0.115$, where it is unstable without the trap.
Hence, the parabolic potential offers (which is also true in the
2D and 3D cases) an additional mechanism of stabilization of the
discrete Skyrmions.

\textit{Conclusions}. We have constructed discrete structures that
emulate Skyrmions on nonlinear dynamical lattices. We have used
the ``hedgehog" ansatz for 2D (\textit{baby}-) Skyrmions in
continuum field models as a guide towards constructing the lattice
solutions and extending them to both 3D and 1D settings. Generally,
the lattice Skyrmion is built herein as a vortex soliton in a complex
field coupled to a dark radial soliton in a coupled real field. We
have predicted stability limits of the lattice Skyrmions and
identified the principal mechanism of their destabilization. The
analytical prediction was verified by numerical computations, that
corroborate the \emph{stability} of the Skyrmions in the predicted
region. We have also demonstrated that the evolution of unstable
discrete Skyrmions leads to onset of lattice turbulence. A
possibility of further stabilization of the Skyrmions by means of
an external confining potential was highlighted too.

It would be 
interesting to examine analogs of the structures presented herein with higher values of the topological
charge, as well as more complex patterns in 
3D. 
Creation of the lattice Skyrmions in binary BECs mixtures
trapped in a deep optical lattice by means of available
techniques appears to be within experimental reach.


\begin{thebibliography}{99}
\bibitem{review} S.\ Aubry, \newblock Physica \textbf{103D}, 201 (1997); S.\
Flach and C. R.\ Willis, \newblock Phys.\ Rep.\ \textbf{295}, 181 (1998);
D. K.\ Campbell {\it et al.}, 
Phys. Today, January
2004, p. 43.

\bibitem{mora} 
A.A.\ Sukhorukov {\it et al.},
IEEE J. Quant. Elect. \textbf{39}, 31 (2003); U. Peschel
\textit{et al.}, J.\ Opt.\ Soc.\ Am.\ B \textbf{19}, 2637 (2002).

\bibitem{PhotCryst} S. F.\ Mingaleev and Y. S.\ Kivshar, Phys.\ Rev.\ Lett.\ 
\textbf{86}, 5474 (2001).

\bibitem{tromb} A.\ Trombettoni and A.\ Smerzi, Phys.\ Rev.\ Lett.\ \textbf{86},
2353 (2001); F. Kh.\ Abdullaev \textit{et al.}, Phys.\ Rev.\
\textbf{A64}, 043606 (2001); F. S.\ Cataliotti \textit{et al.},
Science \textbf{\ 293}, 843 (2001); A.\ Smerzi \textit{et al.},
Phys.\ Rev.\ Lett.\ \textbf{89}, 170402 (2002); G. L.\ Alfimov \textit{et al.},
Phys.\ Rev.\ E \textbf{66}, 046608 (2002); R.\ Carretero-Gonz\'{a}lez and
K.\ Promislow, Phys.\ Rev.\ A \textbf{66} 033610 (2002).


\bibitem{alex} P.\ Binder \textit{et al.}, Phys.\ Rev.\ Lett.\ \textbf{84},
745 (2000); E.\ Tr{\'{\i}}as \textit{et al.},
Phys.\ Rev.\ Lett.\ \textbf{84}, 741 (2000).

\bibitem{6} D. N. Christodoulides \textit{et al.}, 
Nature {\bf 424}, 817 (2003); P. G. Kevrekidis \textit{et
al.}, Int. J. Mod. Phys. B \textbf{15}, 2833 (2001).

\bibitem{7} H. S. Eisenberg {\it et al.}, Phys. Rev. Lett. {\bf 81}, 3383 
(1998); R. Morandotti {\it et al.}, Phys. Rev. Lett. {\bf 86}, 3296 (1999).

\bibitem{8} J. W. Fleischer {\it et al.}, Phys. Rev. Lett. {\bf 90}, 
023902 (2003);
Nature {\bf 422}, 147 (2003).

\bibitem{9} D. Neshev {\it et al.}, Opt. Lett. {\bf 28}, 710 (2003).

\bibitem{10} H. Martin {\it et al.}, Phys. Rev. Lett. {\bf 92}, 123902 (2004).

\bibitem{10a} B.A. Malomed and P. G. Kevrekidis, Phys. Rev. E, \textbf{64},
026601 (2001); B. B. Baizakov {\it et al.}, 
Europhys. Lett. \textbf{63}, 642 (2003); J. Yang and Z.
Musslimani, Opt. Lett. \textbf{28}, 2094 (2003).

\bibitem{11} D. N. Neshev {\it et al.}, Phys. Rev. Lett. {\bf 92}, 
123903 (2004).

\bibitem{12} J. W. Fleischer {\it et al.}, Phys. Rev. Lett. {\bf 92}, 
123904 (2004).

\bibitem{13} J. Yang {\it et al.}, Opt. Lett. {\bf 29}, 1662 (2004).

\bibitem{13a} J. Meier \textit{et al.}, Phys. Rev. Lett. \textbf{91}, 143907
(2003).

\bibitem{14} Z. Chen {\it et al.}, Phys. Rev. Lett. {\bf 92}, 143902 (2004).

\bibitem{14a} J. Yang \textit{et al.}, Phys. Rev. Lett. \textbf{94}, 113902
(2005).

\bibitem{gap} B. Eiermann \textit{et al.}, Phys. Rev. Lett. \textbf{92},
230401 (2004).

\bibitem{skyrme} T.~Skyrme, Proc.\ Roy. Soc.\ London A \textbf{260}, 127
(1961).

\bibitem{hightc} F.~Wilczek, \textit{Fractional Statistics and 
Anyon Superconductivity}, World Scientific, Singapore, 1990.

\bibitem{qhf} D.-H.~Lee and C.~L.~Kane, Phys. Rev. Lett. {\bf 64}, 1313 (1990);
S.~L.~Sondhi \textit{et al.}, Phys. Rev. B \textbf{47}, 16419 (1993).

\bibitem{becskyrme} J. Ruostekoski and J. R. Anglin, Phys. Rev. Lett.
\textbf{86}, 3934 (2001); U. Al. Khawaja and H. T. C. Stoof, Nature (London)
\textbf{411}, 918 (2001); R. A. Battye \textit{et al.}, Phys. Rev. Lett.
\textbf{88}, 80501 (2002); C. M. Savage and J. Ruostekoski, Phys. Rev. Lett.
\textbf{91}, 010403 (2003).

\bibitem{ward} R.~S.~Ward, Lett. Math. Phys. \textbf{35}, 385 (1995).

\bibitem{seibold} G. Seibold, Phys. Rev. B \textbf{58}, 15520 (1998).

\bibitem{elspin} C.~Timm {\it et al.}, 
Phys. Rev. B \textbf{58}, 
10634 (1998).

\bibitem{schramm} A.~J.~Schramm and B.~Svetitsky, Phys. Rev. D \textbf{62}, 
114020 (2000).



\bibitem{photons} 
J. E.\ Heebner and R. W.\ Boyd, J.\ Mod.\ Opt.\ \textbf{49}, 2629 (2002);
P.\ Chak \textit{et al.}, 
Opt.\ Lett.\ \textbf{28}, 1966 (2003). 

\bibitem{weidig} T. Weidig, hep-th/9911056.

\bibitem{piette} A. Kudryavtsev \textit{et al.}, Nonlinearity \textbf{11},
783 (1998).


\bibitem{hall} D. S. Hall \textit{et al.}, Phys. Rev. Lett. \textbf{81},
1539 (1998).




\bibitem{pelin} D.E. Pelinovsky \textit{et al.}, Phys. D \textbf{212}, 20
(2005).

\bibitem{tlm} S. Darmanyan \textit{et al.}, Sov. Phys. JETP \textbf{86}, 682
(1998).

\bibitem{krein} T. Kapitula \textit{et al.}, Physica D \textbf{195}, 263
(2004).

\bibitem{we} P. G. Kevrekidis \textit{et al}., Phys. Rev. Lett. \textbf{93}, 080403 (2004).
\end{thebibliography}
\end{document}